\documentstyle[12pt,aasms4]{article}

\begin{document}

\title{Non-LTE Abundances and Consequences for the Evolution of the
$\rm \alpha$-elements in the Galaxy}
\author{T. Idiart, F. Th\'evenin}
\affil{Universidade de S\~ao Paulo, IAG, Depto. de Astronomia \\
Av. Miguel Stefano 4200, S\~ao Paulo 01065-970, Brazil}
\affil{Observatoire de la C\^{o}te d'Azur \\ 
B.P. 4229, 06304 Nice Cedex 4, France}
\authoremail{thais@iagusp.usp.br,thevenin@obs-nice.fr}

\begin{abstract}
 Abundances of $\alpha$-elements such as Ca and Mg in disk and halo stars 
are usually
derived from equivalent widths lines measured on high resolution spectra, 
and assuming Local Thermodynamic Equilibrium (LTE) .
 In this paper, we present non-LTE differential abundances derived by computing 
the 
statistical equilibrium of CaI and MgI atoms, using high resolution equivalent 
widths available in the literature for 252 dwarf to subgiant stars. 
These non-LTE abundances combined with recent determination of non-LTE 
abundances of iron, 
seem to remove the dispersion of the [Ca/Fe] and [Mg/Fe] ratios in the galactic 
halo and  
disk phases, revealing new and surprising structures. 
These results have important consequences for chemical evolution models of 
the Galaxy. In addition, non-LTE abundance ratios for stars belonging to the M92 
cluster 
apparently have the same behavior. More high resolution observations, 
mainly of globular clusters, are urgently needed to confirm our results.
\end{abstract}


\section{Introduction}

The determination of abundances of nuclear species at distinct locations 
in the Galaxy (e.g. halo, disk and bulge) comes mainly from the spectra of 
late-type star atmospheres. Measured abundances in cool stars at different
stages of evolution give not only the understanding of stellar nucleosynthesis,
but also provide valuable information about the process of chemical
enrichment of the Galaxy.

The archaeological tracers of the chemical evolution
of a star system are the elements produced by explosive nucleosynthesis in
type II (SNII) and type Ia (SNIa) supernovae events. The interest
using such elements as tracers rests on the fact that SNII and SNIa progenitors 
have different lifetimes; SNII is the final evolution of massive
stars and SNIa is a possible final result of evolution of a close binary
system of intermediate mass stars. SNII contribute to the enrichment of the
interstellar medium (ISM) mainly with elements produced by the capture of 
$\alpha$
particles ($\alpha$ elements) and from the r-process, and SNIa produce elements 
belonging to the Fe peak. Consequently, the basic tools to constraint the 
evolution of ISM
in the Galaxy are usually the analysis of relations between ratios of heavy
elements [element/Fe] and Fe abundance [Fe/H].\footnote{$\rm [Fe/H]=
log(N_{Fe}/N_H)-log(N_{Fe}/N_H)_{\odot}$.}

A first glance at the temporal behavior of $\alpha$ elements shows that the 
ratio [$\alpha$/Fe] is
approximately constant for halo metal-poor stars ([Fe/H]$\leq$-1.5)
and decreases for metal-rich stars ([Fe/H]$>$-1.5) belonging to the disk. 
This is reasonably explained by the chemical evolutionary models that assume  
progressive enrichment of ISM by supernovae: first generation of 
stars have in their atmospheres the signature of SNII events only
(called halo-phase of the Galaxy) and the subsequent generations 
have a signature of both SNII and SNIa events (disk-phase).

However, a more precise analysis of [$\alpha$/Fe] vs. [Fe/H] shows 
a pronounced scatter, mainly in the region of metal-poor stars. 
This scatter has been interpreted mostly as a consequence of the inhomogeneity 
of
the matter having made stars rather than resulting of poor observational data 
(Audouze and Silk 1996). 

The derivation of abundances  
based on the analysis of high resolution stellar spectra 
is usually made under the assumption of Local Thermodynamical Equilibrium (LTE).
In the last 15 years, many efforts to estimate errors on abundance 
determinations 
caused by LTE assumption have been done. Recently results for Ba II (Gigas 1986, 
1988,
Mashonkina \& Bikmaev 1996), Sr II (Belyakova \& Mashonkina 1997), Na I 
(Mashonkina et al. 1993), Mg I (Gigas 1986, Mashonkina et al. 1996), Ca I 
(Drake 1991), B I (Kiselman \& Carlsson 1996), Al I (Baumuller \& Gehren 1997) 
and Fe I 
and Fe II (Th\'evenin \& Idiart 1999, TI99), O I (Mishenina et al. 1999) 
and Mg I (Zhao, Butler \& Gehren 1998) demonstrate that most lines can formed
far from LTE conditions. So, some important questions arise: 
what is the influence of non-LTE abundance calculations on the chemical 
evolution 
diagrams of the Galaxy? Do these computations add another different 
constraints to the chemical history of enrichment of the matter in the Galaxy? 

In this work, we present non-LTE abundances derived from computation of 
statistical equilibrium of Ca and Mg atoms, using published equivalent
widths (Sect. 3). The atomic data
and stellar atmospheric models used are presented in Sect. 2.
$\alpha$ elements like Mg and Ca have well-known enhanced abundances 
in atmospheres of F-G metal-poor dwarf stars as a result of cumulative stellar 
generations. Recently, Nissen \& Schuster(1997) and Jehin et al. (1999) proposed 
the 
existence of two sequences of stars having two different [$\alpha$/Fe]
ratios for intermediate stars ([Fe/H] $\approx$ -1). Their works are 
based
on highly accurate observations of stars having approximately same temperatures
and surface gravities.
Based on our non-LTE computations, we found 
different branches or sub-populations of stars not only for intermediate 
metallicities; consequences 
for chemical evolution models of the Galaxy are presented in Sect. 4.
We drawn our conclusions in Sect. 5.

\section{Atomic data and stellar atmospheric models}

The code we used to solve the equations of statistical
equilibrium and radiative transfer is the 2.2 version (1995) of MULTI 
(Carlsson 1986). This code allows us to obtain theoretical spectra of Mg I
and Ca I for given stellar atmospheric models and atomic data.  

The atomic models used are shown by Grotrian diagrams 
presented in Figs. 1 and 2, for Mg I and
Ca I respectively. We included all the fine structure levels of Mg I below 6eV: 
103 levels + continuum and 980 radiative transitions. For calcium, the model has 
83 levels + continuum and 483 radiative transitions.
We used the atomic energy level tables given by Hirata \& Horagushi (1995) 
(HH95)
and Martin et al. (1985). Oscillator strengths are from HH95, Kurucz (1993)
and Th\'evenin (1989, 1990).
We followed the procedure described in TI99 for the remaining
atomic data: radiative and collisional damping 
coefficients, excitation and ionization collisional cross-sections.
Photoionization cross-sections are from TopBase.
The van der Waals damping for all Mg lines was calculated using the 
approximation given by Uns\"old (1955) and multiplied 
by a factor 1.3 as in TI99. For Ca we followed Cayrel et al. (1996).
Note that the factor 1.3 is only important for strong saturated lines. As
will be seen in sect. 3, we always tried to use lines 
lying on the linear part of the curve of growth, where this factor does not play
an important role.      

Stellar atmospheric models are generated using the Gustafsson et al. (1975) and 
Bell 
et al.'s grids (1976),
in order to be consistent with TI99, using $\rm T_{eff}$ from 
Th\'evenin (1998) and $\rm log g$ and [Fe/H] corrected for non-LTE effects 
from TI99. We also 
estimated non-LTE $\rm log g$ and [Fe/H] values for an additional sample of 
stars not analyzed in TI99 (see Table 1). Some objects, to which more accurate 
equivalent widths were recently available, non-LTE surface gravities and [Fe/H]
were re-estimated (see Table 1). 
We emphasize that we do not intend to obtain absolute abundance values 
(see sect. 3), thus in a first approximation we can use LTE atmospheric 
models and perform our analysis just estimating non-LTE effects in statistical 
equilibrium.

\section{Non-LTE calculation and results}

To estimate Ca and Mg non-LTE abundances we used published equivalent widths 
(EW) for 252 stars, including dwarfs, subdwarfs and some subgiants.
Sources of EW data are listed in Table 2 with the respective wavelengths of 
lines used in our analysis. Since we chose to work with one or two lines
at maximum, we selected unsaturated and not very weak lines in order to have a 
more
precise abundance determination.

To obtain non-LTE abundances 
we iterate MULTI with different abundance values until we reproduce 
the measured EW. This kind of procedure has two basic
problems: 1) Our atomic model is not perfect, since, for example we do not 
take into account all the levels and line transitions of all spectroscopic 
terms and certainly there are 
uncertainties in oscillator strengths and in collisional processes with 
neutral H and He (see TI99, for a discussion).
2) Observational errors in EW measurements and inhomogeneity of 
EW from different sources (e.g observations with different
instruments and different placement of the continua).   
  
In order to avoid systematic effects between EW existing in the literature we 
renormalized our resulting abundances using common stars (comparison 
objects) present in the data set of different authors. Of course, a standard 
procedure of renormalization requires
a choice of comparison objects according to each spectral type of the 
analyzed stars. 
In our case this step was not necessary, since our selected 
stars have roughly the same surface 
gravities (dwarfs and subgiants) and a narrow range of not very cool effective 
temperatures (5300 $<$ $T_{eff}$ $<$ 6300). 

All the stars were renormalized to a given reference system, referred here as
Ed93 system (which is based on abundances derived from published EW set of
Edvardsson et al. 1993).
First, we recalculate solar Ca, Mg and Fe abundances using MULTI and 
EW taken from Ed93. These values are our solar references.
Then we estimate the relative abundances [Ca/H] and [Mg/H] for Ed93 stars.
In principle, this differential 
procedure also allows us to minimize the imperfections of the atomic models 
used here. To renormalize the data from other sources we used
observed common stars, as mentioned in the preceding paragraph. For example, for
Mg abundance and mean metallicity $<[Fe/H]>\approx - 1$ 
to include the data of Jehin et al. (J99) into the Ed93 system, nine common 
stars between these two data sets were taken, and the average of the 
differences between estimated abundances was calculated. This averaging 
procedure was used to 
transform J99 data to the Ed93
system. The same procedure was performed on data given by Nissen \& Schuster 
(NS97). For 
metal-poor stars more steps were performed.
For the Zhao \& Magain (ZM90) data, for example, we renormalized
ZM90 into J99 (1 common star), and ZM90 into NS97 (2 common stars), then, as 
J99 and NS97 have common stars with the Ed93 system, these 2 sequences
gave us a renormalization ZM90-Ed93.
 
The main goal of this procedure is to have a homogeneous table of derived 
non-LTE abundances. Results are presented in Table 3 for Ca and Mg.
Fig. 3 shows the abundance ratios [Ca/Fe] and [Mg/Fe] for halo and disk 
phases, presenting parallel structures in both evolutive diagrams. 

Errors are hard to estimate because they involve many factors as described
previously. We estimated the uncertainties in our abundance 
determinations caused by the scatter on derived abundances when varying 
different parameters: stellar (Teff, logg and
[Fe/H]) and observational (EW).
Fig. 4 shows the percentage errors on log($\rm N_{Ca}$) and log($\rm N_{Mg}$)
abundances, corresponding to an incertitude in a given atmospheric 
parameter. These error estimates are made for a star of $\rm \theta_{eff}$=0.85,
logg=4.2 and [Fe/H]=-2.0, for different abundance ratios [Ca/Fe] and [Mg/Fe], 
for CaI and MgI lines of table 2. We see that abundance variations 
are, on the average, more 
sensitive to uncertainties in $\rm T_{eff}$, reaching a maximum of 2.2\% for 
some CaI lines. 
We note that in the case of $\rm log g$ and [Fe/H], the scatters adopted are the 
classical LTE errors, so the abundance errors can be overestimated. 

A similar analysis is made for EW uncertainties using 
the same Ca I and Mg I lines. Fig. 5 shows the percentage errors in EW if we 
have errors in abundance ratios of the order of the structure separations ($\rm 
\approx$ 0.15, 0.2 dex in halo phase) displayed in Fig. 3. For example, the
EW of CaI $\rm \lambda \lambda$ 4578.56 and MgI $\rm \lambda \lambda$ 4571.1 
lines from Zhao \& Magain 1990 have respectively EW errors of 40\% 
and 75\% when [Ca/Fe] and [Mg/Fe] have variations between 0.15 and 0.2 dex. 
The magnitude of these uncertainties
are much greater than the observational errors estimated by EW sources, 
demonstrating that these structures can be real.

Fig. 3 confirms 
that effects of data renormalization as sources of these structures can be
discarded, since the data of distinct sources are distributed 
in different parallel structures. 
As mentioned above, scattering in abundances is sensitive to temperature 
variations,
however this effect seems to be not strong enough to form these structures, as
can be seen in Fig. 6 that shows no correlation between abundance ratios and 
temperatures.

One should have also to keep in mind that such uncertainties in $\rm T_{eff}$ or 
$\rm log g$
result in variations on iron abundances of the same magnitude and sign as
for Ca and Mg abundances, which minimize uncertainties in the abundance 
ratios. Thus, one can say that
the ratios [Ca/Fe] and [Mg/Fe] are less sensitive to these uncertainties on
the stellar fundamental parameters.

\section{Discussion}

The two non-LTE chemical diagrams in Fig. 3 show up that 
the chemical enrichment of the matter in the Galaxy may 
be not so simple as adopted today. On [Ca/Fe] diagram, parallel structures 
appear:
a) halo phase shows three well defined structures separated by 
$\rm \approx $ 0.15 dex and some Ca very deficient stars.
b) in the disk phase (-0.7 $<$ [Fe/H] $<$ 0.4) one can distinguish 
at least 4 structures. This behavior is also present in the intermediate phase 
of the Galaxy (-1.2 $<$ [Fe/H] $<$ -0.7), but only on 3 incurved structures. 

The [Mg/Fe] diagram shows the same structures but with a greater scatter.
This scatter can be due essentially for two reasons.
The first one, because measured EW for magnesium lines are more 
imprecise (there are fewer clean weak lines in observed spectra) 
than for calcium lines and the second, because there are some differences
between the mechanisms of Mg and Ca yields production by SNII
of distinct progenitor masses.
Fig. 7 shows a pronounced scatter in [Mg/H] and [Ca/H] relation in the halo
phase. 

Clearly, the structures pointed out by Nissen $\&$ Schuster (1997)
and Jehin et al.(1999) are present in our Calcium and Magnesium
diagrams, and now extend to the disk and the halo phases of the Galaxy.
Flat structures of $\rm [\alpha/H]$ ratios vs [Fe/H]
are consistent with the idea that both elements come from massive supernovae.
The question is: how field star formation can produce these
structures if, at first, halo stars were formed independently in all the 
protogalaxy phase?
 
The existence of such structures and dichotomy exist also in Globular
Clusters, as it has been demonstrated since more than 15 years. Calcium branch
dichotomy in $\omega$ Cen revealed possible self-enrichment due to SN II,
postulating two epochs of star formation separated by a hiatus time.
Detailed discussion of Norris, Freeman, $\&$ Mighell (1996)
rejects with convincing arguments a merger origin of the dichotomy of [Ca/H] 
abundances
in globular clusters.
Another intriguing coincidence is the bimodality
of C and N abundances on the main sequence of 47 Tuc (Cannon et al. 1998).
As suggested by the authors, an hybrid accretion-enrichment model could
help to understand how globular clusters form and the role played by
stellar winds and those of SN II. Recently Boesgaard et al. (1998) have
observed six turn-off stars in M92, an old very metal-poor cluster. Keeping
in mind that the quality of the spectra is low, giving large uncertainties
to possible derived chemical abundances from then, we derived non-LTE
abundances for iron, calcium and magnesium. As an exercise, we 
plotted the Ca and Mg ratios for these six stars on our diagram
and discovered after renormalisation that they lie exactly at the same place 
as halo stars.
To conclude that this is more than a coincidence is premature but raises an
interesting possibility: a common origin of field and globular cluster stars.
If such observations could be repeated in other globular clusters
for main sequence stars with better signal-to-noise ratios, then they would 
probably help greatly in the understanding of the formation of the Galaxy.
Possible consequences could also be derived concerning the first stars
in the protogalaxy as discussed by Cayrel (1986).

Another possible scenario of the formation of the Galaxy that can produce these
 structures
in the halo phase is an incomplete mixing of SNII yields, as suggested by 
Karlsson \& Gustafsson
(1999). In an even more recent paper, Argast et al. (1999) have explored this 
scenario 
in more detail. Surprisingly, the model proposed by Argast et al.
produces structures in the
[Ca/Fe] and [Mg/Fe] diagrams ([Fe/H] $<$ -1.5) similar to ours, as shown in Fig. 
3. 
Their theoretical 
[Mg/Fe] diagram also shows greater scatter than the [Ca/Fe] diagram, suggesting
 that the origin
of this difference is mainly the yield production mechanism 
(see Argast et al. 1999 for more details), as mentioned in the second paragraph
 of this section. 

For the disk phase, observed structures have a smaller separation, and are more
 evident
in the [Ca/Fe] diagram than in the [Mg/Fe] diagram. One interesting question is: 
how can
chemical evolution
models reproduce this result for disk metal rich stars? Is it that the mixing 
time scale of enriched gas
was larger than the formation time of each generation of stars in the disk, as 
supposed
for the halo phase?  Observations of open cluster stars could verify if they
have similar behavior of globular cluster stars.
 
\section{Conclusions}
We report here non-LTE differential abundances for 252 subdwarf
to subgiant stars using published high resolution equivalent widths.
[Ca/Fe] and [Mg/Fe] diagrams show remarkable 
structures, both in the halo and disk phases of the Galaxy, which are not
related with observational or atmospheric parameter uncertainties. 
These results lead us to a possible evolutive galactic scenario of 
non-homogeneity
or incomplete mixing of synthesized SNII yields (Karlsson \& Gustafsson, 1999, 
Argast et al. 1999). A surprising result is the 
behavior of M92 stars, mainly in the Ca diagram, suggesting a common origin
for field and cluster stars. Spectroscopic high resolution with good S/N 
of stars in clusters are needed to confirm or not the sketch of a new 
chemical evolution model presented in this work. New non-LTE analysis
for other $\alpha$-elements is also necessary to verify if this structural
behavior applies to all $\alpha$-capture SNII products.

\acknowledgements

We thank the referee for many fruitful comments and suggestions
and and for drawing our attention to the paper by Argast et al.
T.I. acknowledges the Brazilian agency FAPESP for the 
grant 97/13083-7 at IAG. This work has been performed using the computing 
facilities provided
by the program Simulations Interactives et Visualisation en Astronomie et 
M\'ecanique (SIVAM)
at the computer center of the Observatoire de la C\^ote d'Azur.

\newpage
\figcaption{Level and Grotrian diagrams of Ca I atom.}

\figcaption{Level and Grotrian diagrams of Mg I atom.}

\figcaption{Chemical evolution diagrams for non-LTE CaI and MgI abundances
estimated in this work and non-LTE [Fe/H] by TI99. Distinct symbols refer 
to EW sources indicated by the legend above. North94, Per86 and Tom95 are
represented with the same symbols (hollow circles)}

\figcaption{Percentage errors $\rm \epsilon$ on CaI and MgI logarithmic
abundances in function of atmospheric parameters variation, for different 
lines of a star with $\rm \theta_{eff}$=0.85, logg=4.2
and [Fe/H]=-2.0. These lines were used for abundance estimates of metal-poor
stars (see table 2). The variation of atmospheric parameters $\rm T_{eff}$, logg 
 and [Fe/H] adopted here are the classical LTE errors, and are on each 
corresponding
diagram. Circles, up and down triangles represent respectively 
[Ca/Fe]=+0.1,+0.25,+0.45 and [Mg/Fe]=+0.6,+0.16,-0.17.}
    
\figcaption{Percentage errors in EW in function of variations of 0.15-0.2 dex 
in the abundance ratios [Ca/Fe]=+0.25 
and [Mg/Fe]=+0.225, for same lines in Fig. 4.}

\figcaption{Chemical abundance ratios in function of effective temperature 
($\rm \theta _{eff}=5040/T_{eff}$). Different symbols follow
the same legend of Fig. 3.}
   
\figcaption{Relation between Mg and Ca abundances. Different symbols follow
the same legend of Fig. 3.}

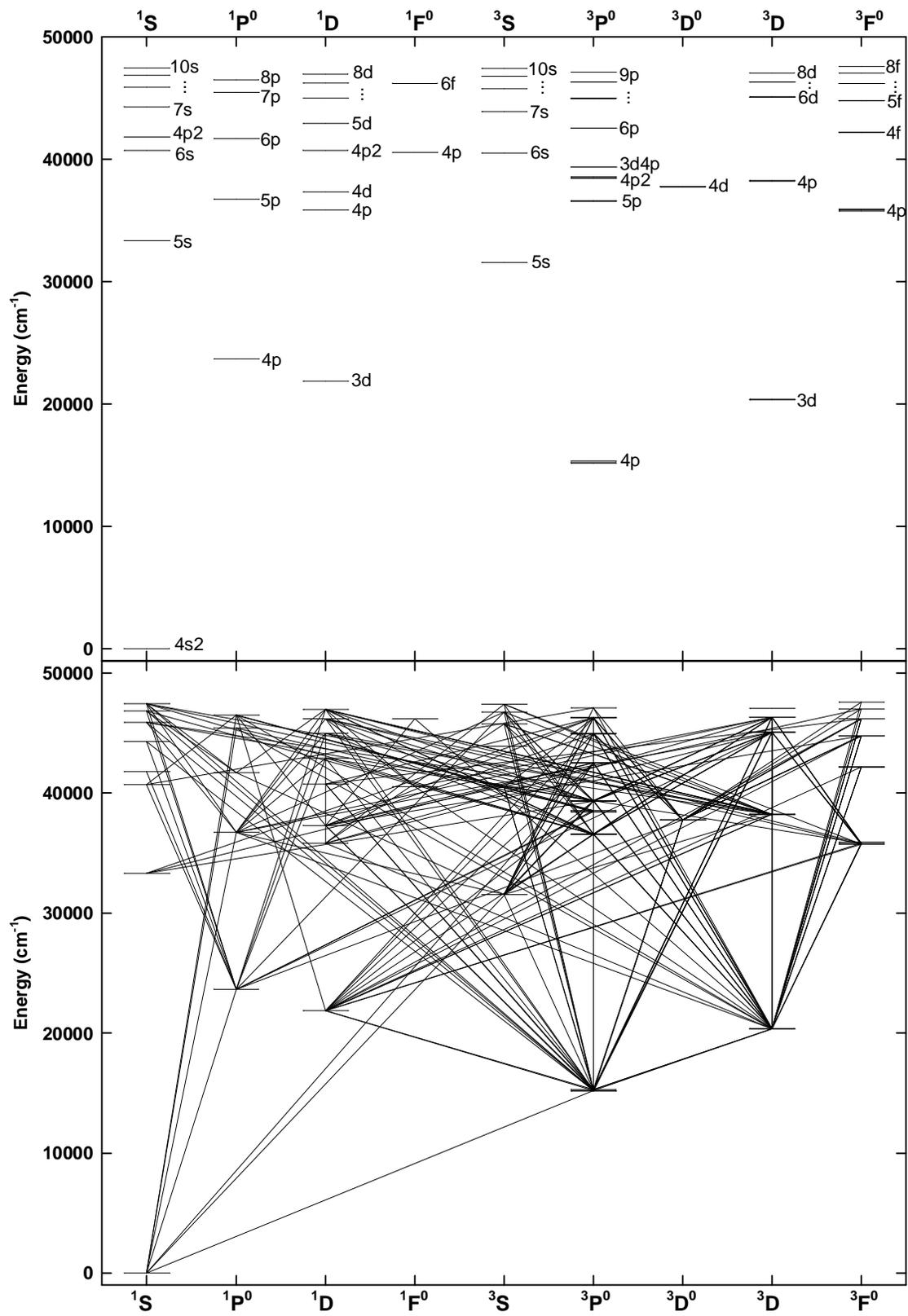
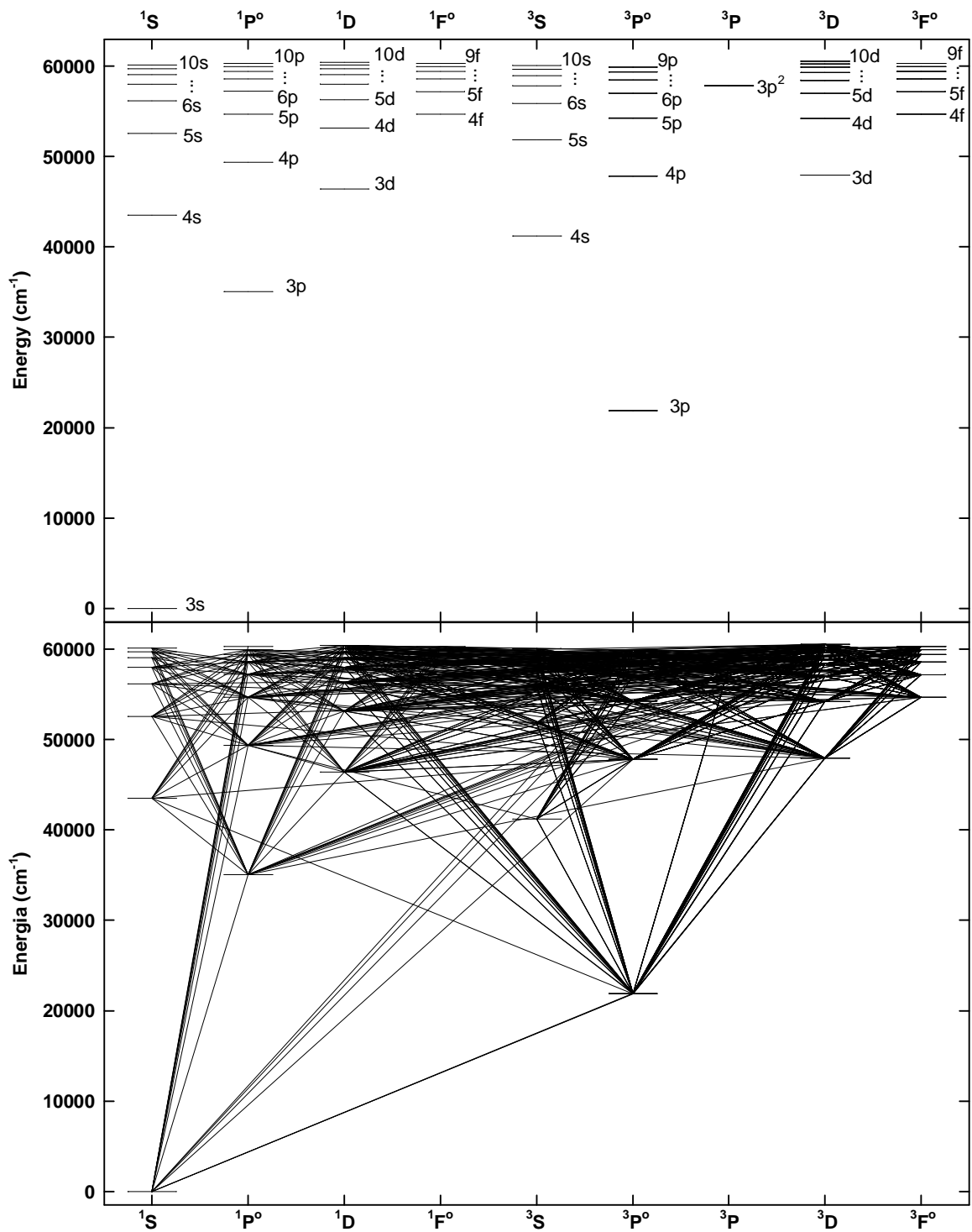
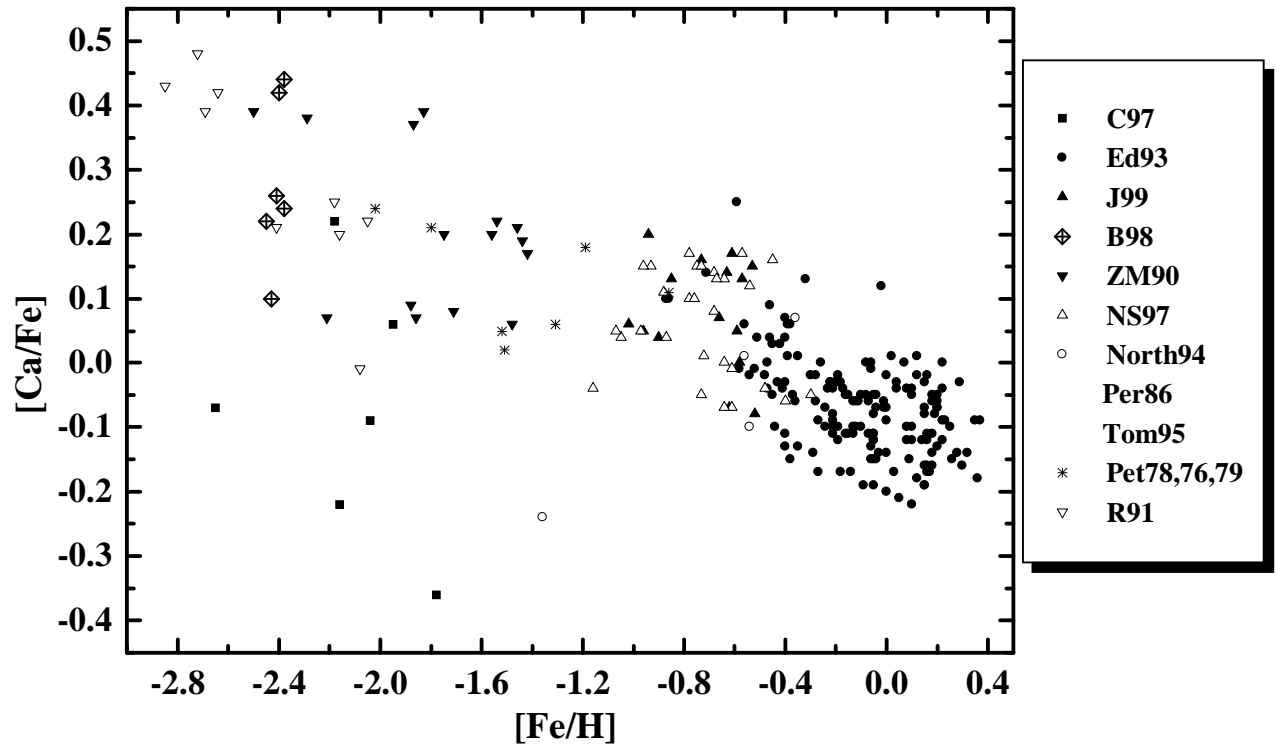
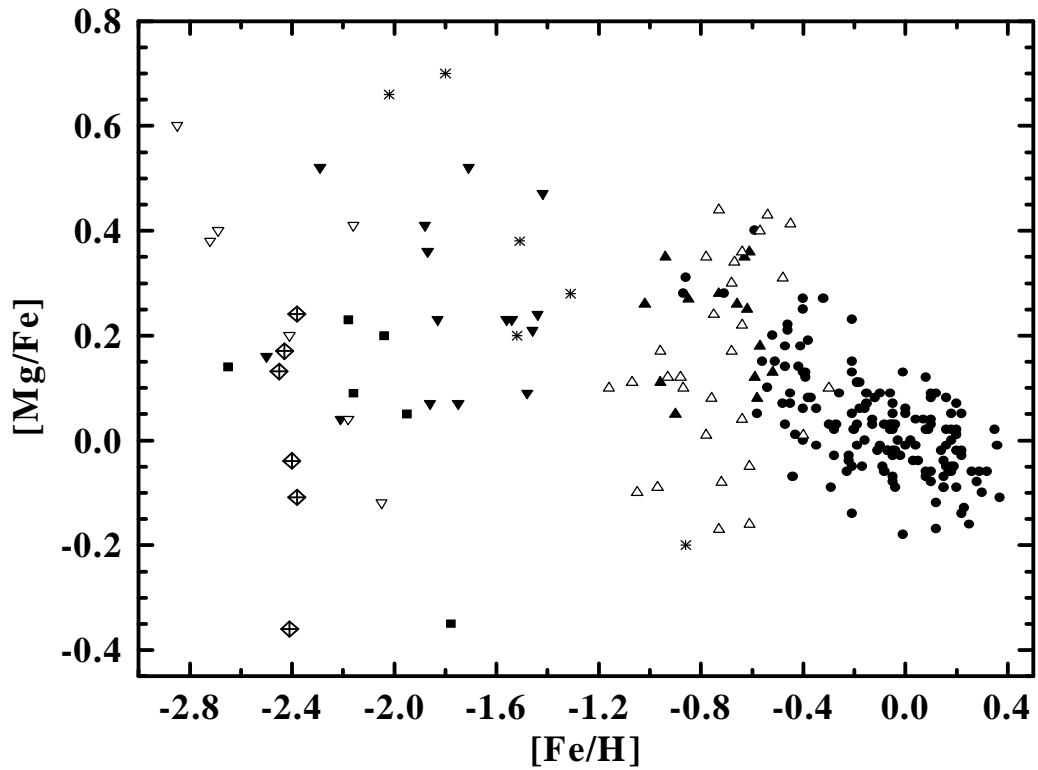
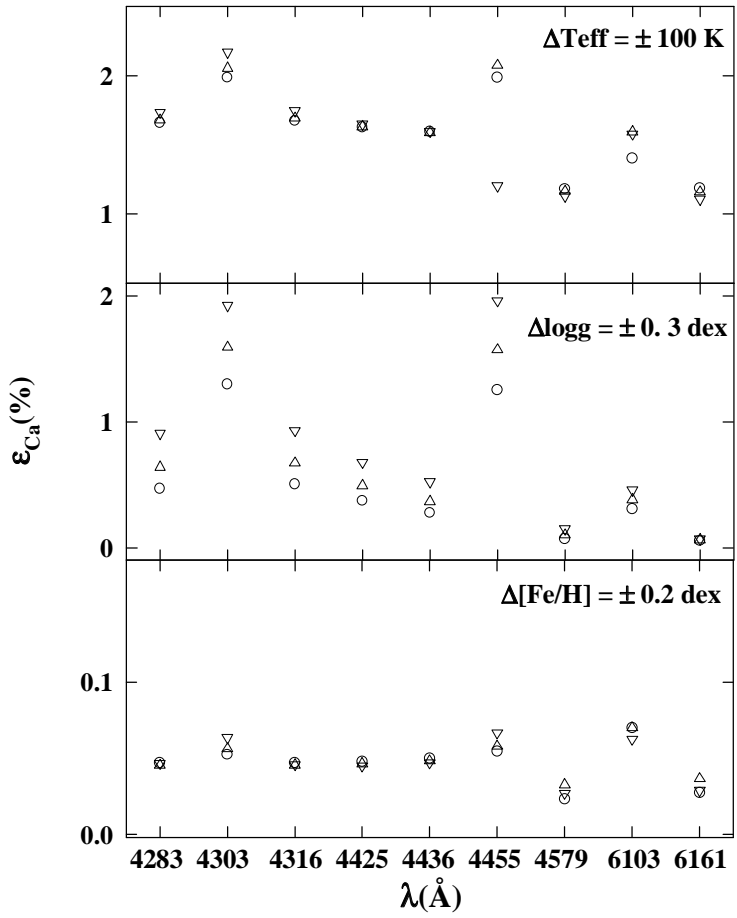
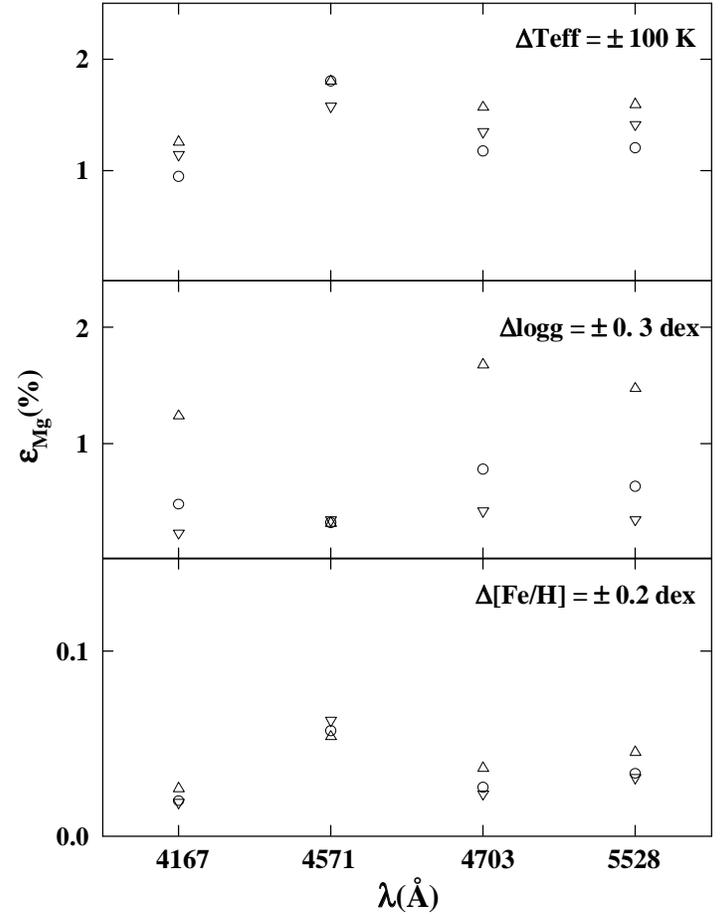
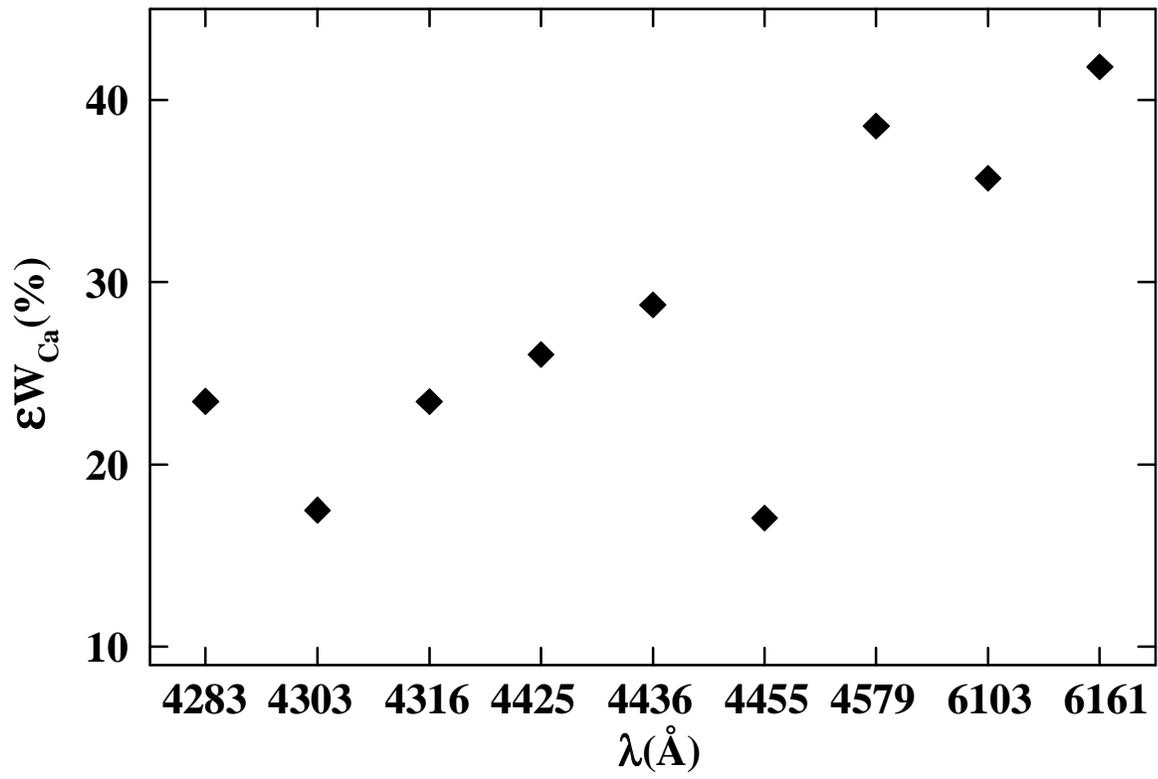
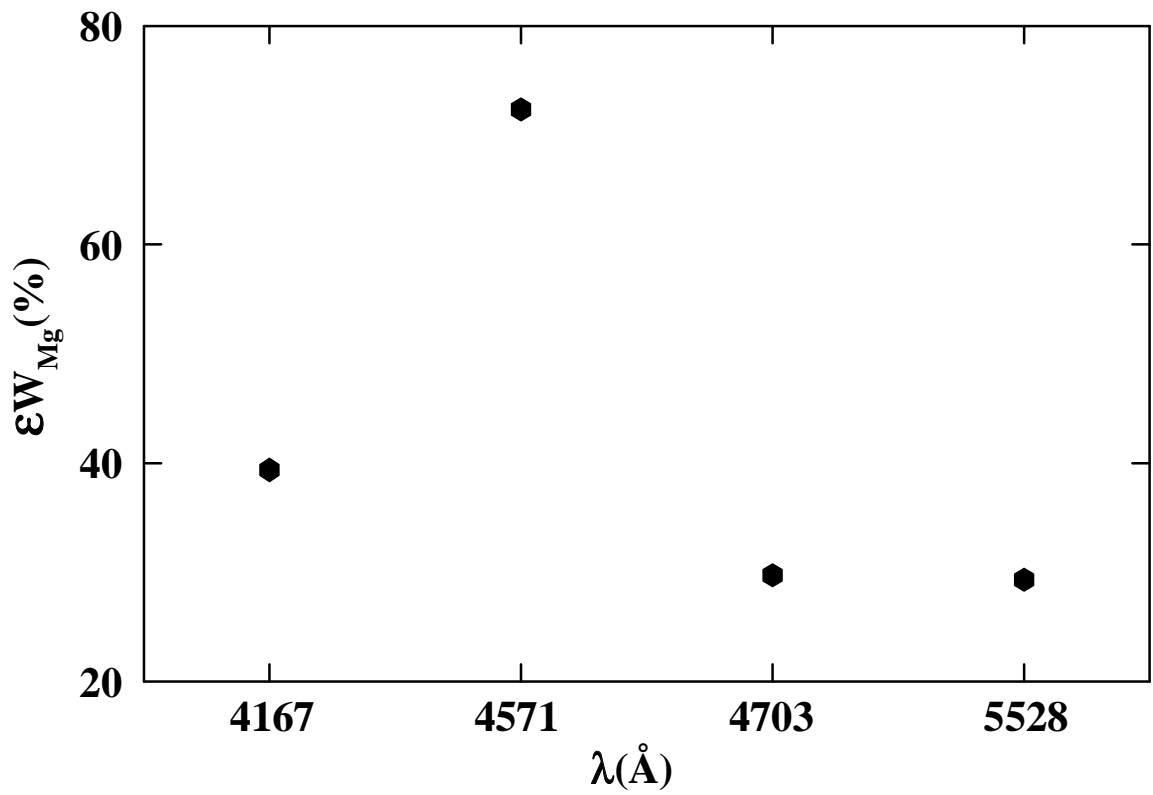
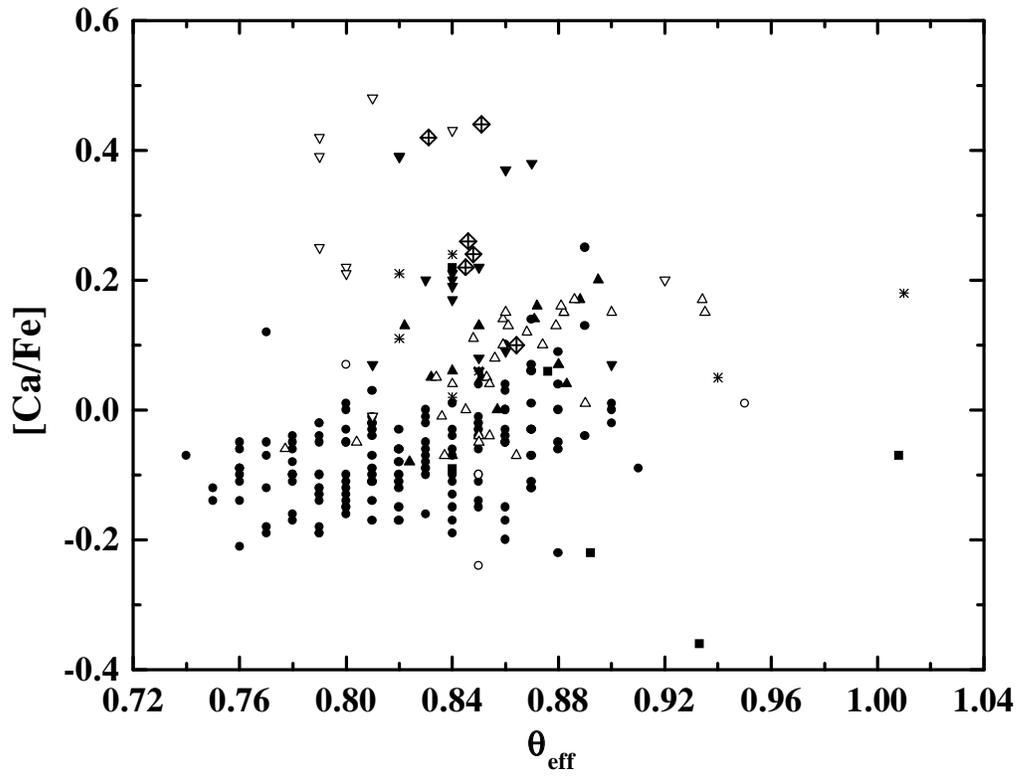
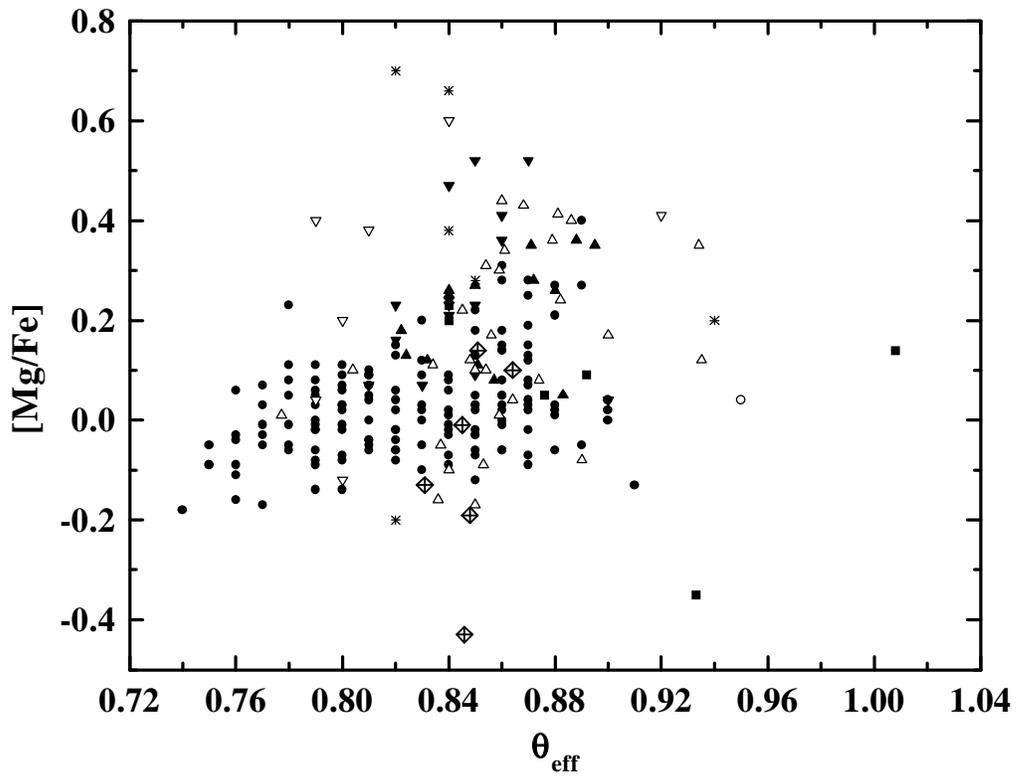
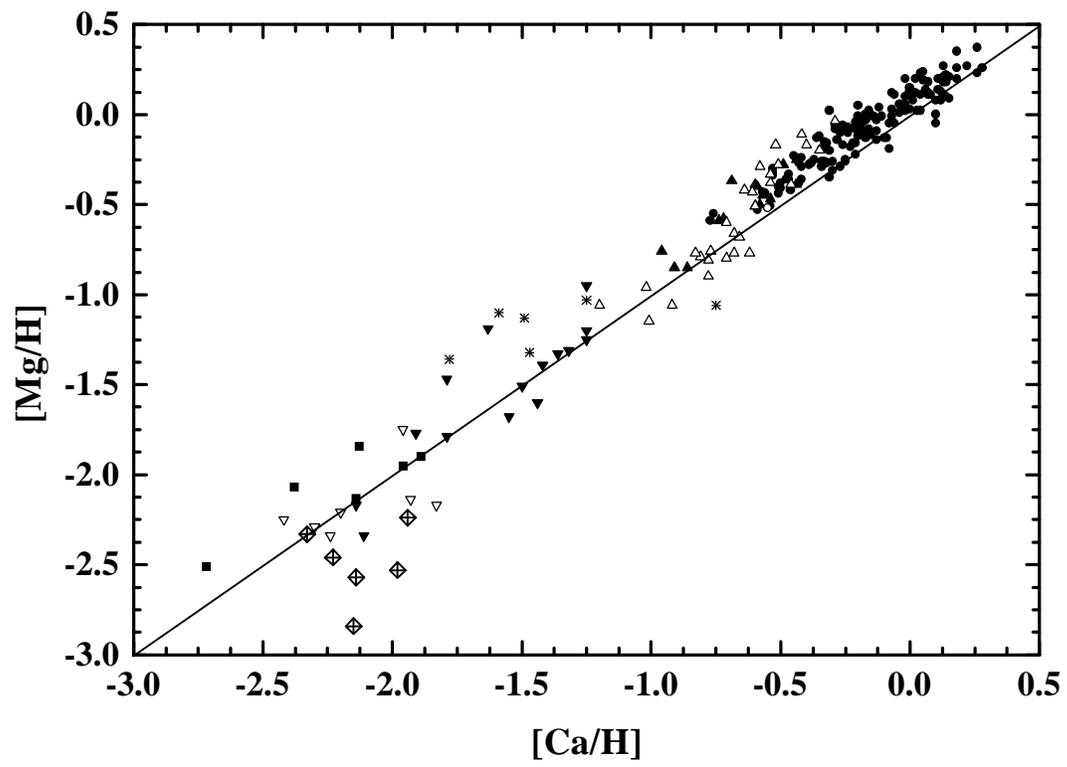
\end{document}